\documentclass[twocolumn]{IEEEtran}
\usepackage{amsmath}
\usepackage{amssymb}
\usepackage{float}
\usepackage{multirow}
\usepackage{graphicx}
\usepackage{color}

\newtheorem{thm}{Theorem}

\newtheorem{rem}{Remark}
\newtheorem{defn}{Definition}

\newtheorem{lem}{Lemma}

\floatstyle{ruled}
\newfloat{algorithm}{tbp}{loa}
\providecommand{\algorithmname}{Algorithm}
\floatname{algorithm}{\protect\algorithmname}
\usepackage{algorithm}
\usepackage{algorithmic}

\begin{document}

\title{A Fast Converging Distributed Solver for Linear Systems with Generalised Diagonal Dominance}
\author{{\normalsize{Qianqian Cai$^1$, Zhaorong Zhang$^2$ and Minyue Fu$^{2,1}$, {\em Fellow, IEEE}}} 
\thanks{$^1$School of Automation, Guangdong University of Technology, and Guangdong Key Laboratory of IoT Information Processing, Guangzhou 510006, China.}
\thanks{$^2$School of Electrical Engineering and Computer Science, The University of Newcastle. University Drive, Callaghan, 2308, NSW, Australia.}
\thanks{This work was supported by the National Natural Science Foundation of China (Grant Nos.~61633014 and U1701264). E-mails: 
qianqian.cai@outlook.com; zhaorong.zhang@uon.edu.au; minyue.fu@newcastle.edu.au.}
}
\maketitle
\begin{abstract}
This paper proposes a new distributed algorithm for solving linear systems associated with a sparse graph under a generalised diagonal dominance assumption. The algorithm runs iteratively on each node of the graph, with low complexities on local information exchange between neighbouring nodes, local computation and local storage. For an acyclic graph under the condition of diagonal dominance, the algorithm is shown to converge to the correct solution in a finite number of iterations, equalling the diameter of the graph. For a loopy graph, the algorithm is shown to converge to the correct solution asymptotically. Simulations verify that the proposed algorithm significantly outperforms the classical Jacobi method and a recent distributed linear system solver based on average consensus and orthogonal projection.   
\end{abstract}


\section{Introduction}

Sparse linear systems are of great interest in many disciplines, and many iterative methods exist for solving sparse linear systems; see, e.g., \cite{Golub,Horn,Saad,Mahmood,Ran,Rozsa,Abur,Khan,Xiao,Weiss,Shental,Mailoutov}. However, for many networked system applications, distributed algorithms are preferred. The main difference is that an iterative method focuses on reducing the total computation by incorporating the sparsity of the system, but allowing a central controller for the coordination of the algorithm. In contrast, a distributed algorithm is designed such that the same algorithm is executed concurrently on every node (or sub-system, or agent) while allowing only neighbouring nodes to share information. Ideally, no central controller or coordinator is required, and each node should have low complexities in terms of information exchange, computation and storage. For example, a sequential iterative algorithm, where the nodes are ordered and their executions are done sequentially, should not be regarded as a distributed algorithm. For a large networked system such as sensor networks \cite{Kar,Fu,Fu1}, networked control systems \cite{Morse,Morse1,Wang}, network-based state estimation~\cite{Damian,Russell,Bertrand,Bertrand1}, biological networks \cite{Vicsek1,Vicsek2}, {multi-agent systems~\cite{Vicsek3,Lin,Fu2,Fu3}, multi-agent optimization~\cite{Nedic,Jak,Lu} and so on, it is often desirable to use distributed algorithms rather than iterative methods.

In this paper, we study distributed solutions for sparse linear systems. Our focus is on linear systems which satisfy the so-called generalised diagonal dominance condition. These systems include those with diagonal dominance as a subset (those with a diagonally dominant matrix). Diagonally dominant matrices have vast applications in engineering, economics and finance \cite{Golub,Horn}. For example, many matrices that arise in finite element modeling are diagonally dominant \cite{Reddy}; diagonally dominant matrices also arise naturally for covariance of partially correlated random vectors \cite{Fu4}.  The generalised diagonal dominance condition relaxes the diagonal dominance requirement using a diagonal scaling matrix, i.e., only a diagonally scaled matrix is required to be diagonally dominant. As we will point out later, the generalised diagonal dominance condition is also equivalent to the so-called walk-summability condition~\cite{Mailoutov,Li}.

We propose a new distributed algorithm for solving sparse linear systems with generalised diagonal dominance. The algorithm runs iteratively on each node of the graph associated with the system, and enjoys low complexities on local information exchange between neighboring nodes, local computation and local storage. For an acyclic graph, the algorithm converges to the exact solution in $d$ iterations, where $d$ is the diameter of the graph. This is also the theoretical minimum number of iterations required to solve a linear system using a distributed algorithm. For a loopy graph, the algorithm is guaranteed to converge to the exact solution asymptotically. 

We have compared the proposed algorithm with the classical Jacobi method \cite{Golub,Horn,Saad} and the distributed linear system solver \cite{Morse} based on average consensus and orthogonal projection to show that our algorithm has much faster convergence. 

The development of the algorithm and its theoretical analysis follows from a similar algorithm known as Gaussian belief propagation (BP) algorithm \cite{Pearl} which applies to Gauss-Markov random fields which involve a symmetric system matrix; see, e.g., \cite{Weiss,Mailoutov}. Namely, the convergence properties in \cite{Weiss} are for symmetric matrices with diagonal dominance, and the convergence properties in \cite{Mailoutov} are for symmetric matrices with generalised diagonal dominance (or walk summability).  The main contribution of our work in comparison with those in \cite{Weiss,Mailoutov} can be viewed as generalising the Gaussian BP algorithm to solving linear systems with an asymmetric system matrix. 

\section{Problem Formulation and Preliminaries}

\subsection{Problem Formulation}
Consider a network of nodes $(1,2,\ldots, n)$ associated with a state vector $x=\mathrm{col}\{x_1,x_2,\ldots, x_n\}\in \mathbb{R}^n$. The information available at each node $i$ is that $x$ satisfies a linear system:
$$a_i^T x = b_i,$$ 
where $a_i=\mathrm{col}\{a_{i1}, a_{i2}, \ldots\, a_{in}\}\in \mathbf{R}^{n}$ is a column vector and $b_i$ is a scalar.  Collectively, the common state satisfies 
\begin{align}
A x &= b \label{Axb}
\end{align}
with $A=\mathrm{col}\{a_1^T, a_2^T, \ldots, a_n^T\}$ and $b=\mathrm{col}\{b_1, b_2, \ldots, b_n\}$.  
It is assumed throughout the paper that (i) $A$ is invertible; and (ii) $a_{ii}\ne0$ for all $i=1,2,\ldots, n$. Note that (ii) can be guaranteed under (i) because nonzero diagonals can always be obtained through proper pivoting (swapping of rows of columns) \cite{Horn}. 
Denote the solution of (\ref{Axb}) by 
\begin{align}
x^{\star}&=A^{-1}b. \label{xstar}
\end{align} 

Define the {\em induced graph} $\mathcal{G}=\{\mathcal{V}, \mathcal{E}\}$ with $\mathcal{V}=\{1, 2, \ldots, n\}$ and $\mathcal{E}=\{(i,j)|a_{ij}\ne 0 \mathrm{\ or\ } a_{ji}\ne0\}$. We associate each node $i$ with $(a_{ii}, b_i)$,  and each edge $(i, j)$ with $a_{ij}$. Note in particular that $\mathcal{G}$ is {\em undirected}, i.e., $(i,j)\in \mathcal{E}$ if and only if $(j,i)\in\mathcal{E}$.  For each node $i\in \mathcal{V}$, we define its {\em neighbouring set} as $\mathcal{N}_i=\{j|(i,j)\in \mathcal{E}\}$ and denote its {\em cardinality} by $|\mathcal{N}_i|$. We assume in the sequel that $|\mathcal{N}_i|\ll n$ for all $i\in \mathcal{V}$. The graph $\mathcal{G}$ is said to be {\em connected} if for any $i, j\in \mathcal{V}$, there exists a (connecting) {\em path} of  $(i, i_1), (i_1, i_2), \ldots, (i_k, j)\in \mathcal{E}$. The {\em distance} between two nodes is the minimum number of edges connecting the two nodes. It is obvious that a graph with a finite number of nodes is either connected or composed of a finite number of disjoint subgraphs with each of them being a connected graph. The {\em diameter} of a connected graph is defined to be largest distance between two nodes in the graph. The diameter of a disconnected graph is the largest diameter of a connected subgraph. By this definition, the diameter of a graph with a finite number of nodes (or a finite graph) is always finite.  A {\em loop} is defined to be a path starting and ending at node $i$ through a node $j\ne i$. A graph is said to be {\em acyclic} if it does not contain any loop. A {\em cyclic} (or {\em loopy}) graph is a graph with at least one loop.  

The {\em distributed linear system problem} we are interested in is {\bf to devise an iterative algorithm for every node $i\in \mathcal{V}$ to execute so that the computed estimate $\hat{x}_i(k)$ at iteration $k=0,1,2,\ldots,$ will approach $x_i^{\star}$} as $k$ increases.  

Note that in this problem formulation, node $i$ is only interested in its local variable, and not the solution of $x_j$ for any other node $j\in \mathcal{V}$. This is {\em sharply different} from distributed methods in the literature (e.g., \cite{Morse}) which require each node to compute the whole solution of $x$. For a large network, computing the whole solution of $x$ is not only burdensome for each node, but also unnecessary in most applications. 

We want the algorithm to be of low complexity and fast convergence.   Certain {\em constraints} need to be imposed on the algorithm's complexities of communication, computation and storage to call it {\em distributed}. In our paper, these include:
\begin{enumerate}
\item[C1:] Local information exchange:  Each node $i$ can exchange information with each $j\in \mathcal{N}_i$ only once per iteration. 
\item[C2:] Local computation: Each node $i$'s computational load should be at most $O(|\mathcal{N}_i|)$ per iteration. 
\item[C3:] Local storage: Each node $i$'s storage should be at most $O(|\mathcal{N}_i|)$ over all iterations. 
\end{enumerate}

\begin{lem}\label{lem:0.1}
Under the local information exchange constraint above, a connected graph $\mathcal{G}$ with diameter $d$ needs a minimum of $d$ iterations to solve the distributed linear system problem with a general $A$ and $b$. 
\end{lem}

\begin{IEEEproof} Let nodes $i$ and $j$ be such that they are $d$ hops away from each other (such nodes exist by the definition of $d$). The correct solution for $x_i$ requires the information of all the $b_j$ connected to node $i$, which needs to propagate to node $i$. By the local information exchange constraint, this will take at least $d$ iterations.  
\end{IEEEproof}
\begin{rem}\label{rem:Axb}
Although we focus only on linear systems with square and invertible matrix $A$, under-determined and over-determined linear systems can be treated with slight modifications. For the under-determined case (where the row rank of $A$ is less than $n$), we can zero-pad $(A,b)$ to make $A$ square, then solve the regularised problem of $(A+\lambda I)x=b$ for some sufficiently small scalar or diagonal matrix $\lambda>0$. For the over-determined case (where $A$ has full column rank and  has more rows than columns), it is standard to solve the least-squares problem $\min_x (Ax-b)^T(Ax-b)$ instead, which results in the linear system $(A^TA)x=(A^Tb)$ with a square and invertible $(A^TA)$. 
\end{rem}  
\begin{rem}\label{rem:Jacobi}
The well-celebrated {\em Jacobi method} (see, e.g., \cite{Golub,Horn,Saad}) for solving a linear system with diagonal dominance is an excellent example of distributed algorithms. Denoting by $\hat{x}_i(k)$ the estimate of the $i$-th component of $x^{\star}$, this method is simply rewritten as 
\begin{align}
\hat{x}_i(k+1) &= \frac{1}{a_{ii}}\left (b_i-\sum_{j\in \mathcal{N}_i} a_{ij}\hat{x}_j(k)\right ) \label{Jac}
\end{align} 
with the initial condition of $x_i(0)=b_i/a_{ii}$, $i=1,2, \ldots, n$. It is straightforward to verify that constraints C1)-C3) are all met by this method. In contrast, consider the well-known enhanced version of the Jacobi method known as the Gauss-Seidel algorithm (also see \cite{Golub,Horn,Saad}):
\begin{align}
\hat{x}_i(k+1) &=\hspace{-1mm} \frac{1}{a_{ii}}\hspace{-1mm}\left (b_i-\hspace{-2mm}\sum_{\stackrel{j<i}{j\in \mathcal{N}_i}} a_{ij}\hat{x}_j(k+1)-\hspace{-1mm}\sum_{\stackrel{j>i}{j\in \mathcal{N}_i}} a_{ij}\hat{x}_j(k)\hspace{-1mm}\right ) \label{Gas}
\end{align} 
This can only be implemented sequentially from $i=1$ to $n$ due to the fact that the information of $x_j(k+1)$, $j<i$, is required for computing $\hat{x}_i(k+1)$. That is, node $2$ needs to wait for node $1$ to finish its iteration, and so on. Because of this, distributed implementation of the Gauss-Seidel algorithm is not possible. A similar comment applies to the more general  successive over relaxation (SOR) method and symmetric SOR (SSOR) method (also see \cite{Golub,Horn,Saad}).
\end{rem}

\subsection{Walk Summability and Generalised diagonal dominance}

The linear systems we are studying in this paper belong to the class of {\em walk-summable systems}.  This represents a large class of systems, including {\em diagonally dominant systems}, which find wide applications in scientific and engineering disciplines, as a sub-class. 

The notion of {\em walk summability} is generalised from Gaussian graphical models \cite{Mailoutov}. Given an $n\times n$ matrix $R=\{r_{ij}\}$ and its induced graph $\mathcal{G}=\{\mathcal{V}, \mathcal{E}\}$, a {\em walk} of length $\ell\ge0$ in $\mathcal{G}$ is a sequence $w=(w_0, w_1, \ldots, w_{\ell})$ of nodes $w_k\in \mathcal{V}$ with $(w_k, w_{k+1})\in \mathcal{G}$ for all $k$.  The {\em weight} of the walk is defined to be 
\begin{align}
\phi(w) &= \prod_{k=1}^{\ell} r_{w_{k-1},w_k} \label{eq:walk}
\end{align}
For a zero-length ``self'' walk $w=(v)$ at node $v$, the convention of $\phi(w)=1$ is used. If $w$ is an empty set, the convention is that $\phi(w)=0$. For notational simplicity, denote by $i\stackrel{\ell}{\mapsto} j$ a length-$\ell$ walk from node $i$ to $j$, and denote by $i\mapsto j$ a walk from node $i$ to $j$ without specifying its length. The $(i,j)$-th element of $R^{\ell}$ can be expressed as a sum of weights of walks (or simply called {\em walk sum}) that go from $i$ to $j$ with length $\ell$, i.e., 
\begin{align}
(R^{\ell})_{ij} = \sum_{w_1,\ldots, w_{\ell}} r_{i,w_1} r_{w_1,w_2}\ldots r_{w_{\ell-1},j} = \sum_{i\stackrel{\ell}{\mapsto} j} \phi(w) \label{eq:walk1}
\end{align}

\begin{defn}\label{def:walk}
A given matrix $R$  is said to be {\em walk-summable} if for all $i,j\in \mathcal{V}$, the unordered sum over all walks from $i$ to $j$ in the induced graph $\mathcal{G}=\{\mathcal{V}, \mathcal{E}\}$, as expressed by $\sum_{w: i\mapsto j} \phi(w)$, is well defined. The linear system (\ref{Axb}) with $a_{ii}=1$ for all $i$ is said to be {\em walk-summable} if $R=I-A$ is walk-summable. 
\end{defn}

Define the walk sum for a set of walks $\mathcal{W}$ to be 
\begin{align*}
\phi(\mathcal{W}) = \sum_{w\in \mathcal{W}} \phi(w) 
\end{align*}
The walk sum enjoys the following basic properties \cite{Mailoutov}:
\begin{itemize}
\item[P1] For two walk sets $\mathcal{U}$ and $\mathcal{V}$, the product set $\mathcal{UV}=\{uv | u\in \mathcal{U}, v\in \mathcal{V}\}$ satisfies $\phi(\mathcal{UV})=\phi(\mathcal{U})\phi(\mathcal{V})$. In particular, for the $\ell$-fold product $\mathcal{U}^{\ell}=\mathcal{U}\ldots \mathcal{U}$, $\phi(\mathcal{U}^{\ell})=\phi(\mathcal{U})^{\ell}$. 
\item[P2] If $\mathcal{W}_k$ are disjoint walk sets for $k=1, 2,\ldots,$ then $\phi(\cup_{k=1}^{\infty} \mathcal{W}_k)=\sum_{k=1}^{\infty} \phi(\mathcal{W}_k)$.  
\item[P3] If $\mathcal{W}_k\subset \mathcal{W}_{k+1}$ for all $k=1,2,\ldots$, then $\phi(\cup_{k=1}^{\infty} \mathcal{W}_k)=\lim_{k\rightarrow\infty} \phi(\mathcal{W}_k)$.
\end{itemize}

Other properties of walk-summability and its relationship with diagonally dominant systems and its equivalence with the so-called generalised diagonally dominant systems can be found in Appendix. 

A particularly important property of walk-summable system is that the corresponding matrix $R=I-D_A^{-1}A$ has spectral radius less than 1 (see Lemma~\ref{lem:walk}), where $D_A=\mathrm{diag}\{A\}$. It follows that \cite{Mailoutov}
\begin{align*}
A^{-1} &=(I-R)^{-1}D_A^{-1} = (\sum_{\ell} R^{\ell}) D_A^{-1}\\
x^{\star} &=A^{-1}b = \sum_{\ell} R^{\ell} (D_A^{-1}b).
\end{align*}
This leads to the following key technical lemma  \cite{Mailoutov}, which reveals the important role of walk-summability in distributed solutions for linear systems.
\begin{lem}\label{lem:key}
Denoting $P=\{p_{ij}\}=(I-R)^{-1}$, it follows that
\begin{align}
p_{ij} &= \sum_{w:i\mapsto j} \phi(w) = \phi(\{i\mapsto j\}), \\
x_i^{\star} &=\sum_{s}P_{is}(b_s/a_{ss})=\sum_{s}\phi(\{s\mapsto i\}) (b_s/a_{ss}). \label{eq:walk_xi}
\end{align}
\end{lem}
In the above, $\{i\mapsto j\}$ denotes the set of all walks for $R$ from node $i$ to node $j$. 

\section{Distributed Solver for Linear systems}

Our proposed algorithm is given in Algorithm~\ref{alg1}. In each iteration $k$ of the algorithm, each node $i$ computes variables $a_{i\rightarrow j}(k)$ and $b_{i\rightarrow j}(k)$ for each of its neighboring node $j\in \mathcal{N}_i$ and transmit them to node $j$.  All the nodes execute the same algorithm concurrently.  

\begin{algorithm}[h] 
\protect\protect\protect\protect\protect\protect\protect\caption{(Distributed Solver for Linear systems)}
\label{alg1} \begin{itemize}

\item \textbf{Initialization:} For each node $i$, do: For each $j\in \mathcal{N}_i$, set $a_{i\rightarrow j}(0) = a_{ii}, b_{i\rightarrow j}(0)=b_i$ and transmit them to node $j$.

\item \textbf{Main loop:} At iteration $k=1,2,\cdots$, for each node $i$, compute
\begin{align}
\tilde{a}_i(k) &= a_{ii} - \sum_{v\in \mathcal{N}_i} \frac{a_{vi}a_{iv}}{a_{v\rightarrow i}(k-1)} \label{ak0}\\
\tilde{b}_i(k) &= b_i-\sum_{v\in \mathcal{N}_i} \frac{a_{iv}b_{v\rightarrow i}(k-1)}{a_{v\rightarrow i}(k-1)}
\label{bk0}\\
\hat{x}_i(k)& = \frac{\tilde{b}_i(k)} {\tilde{a}_i(k)}, \label{xik}
\end{align}
then for each $j\in \mathcal{N}_i$, compute
\begin{align}
a_{i\rightarrow j}(k) &= \tilde{a}_i(k)+ \frac{a_{ji}a_{ij}}{a_{j\rightarrow i}(k-1)} \label{ak}\\
b_{i\rightarrow j}(k) &= \tilde{b}_i(k) + \frac{a_{ij}b_{j\rightarrow i}(k-1)}{a_{j\rightarrow i}(k-1)} \label{bk}
\end{align}
and transmit them to node $j$. 
\end{itemize}
\end{algorithm}

We now provide two main results about Algorithm~\ref{alg1}, one for  acyclic graphs (Theorem~\ref{thm:1}) and one for loopy graphs (Theorem~\ref{thm:2}).  Their proofs will be postponed to the next two sections.  

\begin{thm}\label{thm:1}
Suppose the linear system (\ref{Axb}) is walk-summable (i.e., generalised diagonally dominant) and the induced graph $\mathcal{G}$ is acyclic with diameter $d$. Then,  running Algorithm~\ref{alg1}  gives the following:
\begin{itemize}
\item $a_{i\rightarrow j}(k)>0$ for all $i\in \mathcal{V}, j\in \mathcal{N}_i$ and $k=0,1, \ldots$;
\item $a_{i\rightarrow j}(k)=a_{i\rightarrow j}(d-1)$, $b_{i\rightarrow j}(k)=b_{i\rightarrow j}(d-1)$, $\tilde{a}_i(k)=\tilde{a}_i(d)$ and $\tilde{b}_i(k)=\tilde{b}_i(d)$ for all $i\in \mathcal{V}, j\in \mathcal{N}_i$ and $k\ge d$.
\end{itemize}
More importantly, 
\begin{align}
\hat{x}_i(k) &= x_i^{\star},    \  \ \ \forall\  k\ge d,  \ i\in \mathcal{V}. \label{xid}
\end{align} 
\end{thm}

\begin{thm}\label{thm:2}
Suppose the linear system (\ref{Axb}) is walk-summable (i.e., generalised diagonally dominant). Then, running Algorithm~\ref{alg1}  yields $a_{i\rightarrow j}(k)>0$ for all $i\in \mathcal{V}, j\in \mathcal{N}_i$ and $k=0,1, \ldots$, and that  
\begin{align}
\lim_{k\rightarrow \infty} \hat{x}_i(k)=x_i^{\star}, \ \ \forall \ i\in \mathcal{V}.\label{loop_xik2}
\end{align}
\end{thm}

\section{Convergence Analysis for Acyclic Graphs}

In this section, we consider a tree graph $\mathcal{G}$ and proceed to prove Theorem~\ref{thm:1}. 

Suppose $\mathcal{G}$ is a connected graph with diameter $d$.  Take any node $i\in \mathcal{V}$ and $j\in \mathcal{N}_i$, it is obvious that $\mathcal{G}$ will become two disjoint subgraphs if we remove the edge $(i,\ j)$. Denote by $\mathcal{G}_{i\backslash j}$ the subgraph containing $i$, i.e., it is obtained from $\mathcal{G}$ by removing all the nodes and edges connected to $i$ through $j$ (including node $j$ and the edge $(i,\ j)$). Denote by $\mathcal{G}_{i\backslash j}(k)$ the subgraph of $\mathcal{G}_{i\backslash j}$ containing only the nodes which are within $k$ hops away from node $i$. An example of such a subgraph is $\mathcal{G}_{2\backslash 1}(1)$ in Fig.~\ref{fig:S.1} which contains the nodes 2, 4 and 5 only. In particular, $\mathcal{G}_{i\backslash j}(0)$ is a singleton graph containing only $i$, and $\mathcal{G}_{i\backslash j}(k)=\mathcal{G}_{i\backslash j}(d-1)=\mathcal{G}_{i\backslash j}$ for all $k\ge d$ due to the definition of graph diameter.  Denote by $i\mapsto i \mid \mathcal{G}_{i\backslash j}(k)$ a return walk for node $i$ constrained in $\mathcal{G}_{i\backslash j}(k)$. The set of such walks is denoted by $\{i\mapsto i\} \mid \mathcal{G}_{i\backslash j}(k)$.  For the subgraph $\mathcal{G}_{2\backslash 1}(1)$ in Fig.~\ref{fig:S.1}, an example of such a walk is $w=(2,4,2,5,2,4,2)$.

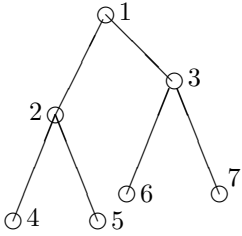
\begin{figure}[ht]
\begin{picture}(240,95)
\put(124,93){\circle{6}}
\put(124,93){\line(1,-1){25}} 
\put(150,68){\circle{6}}
\put(155,66){$3$}
\put(129,91){$1$}
\put(105,55){\circle{6}}
\put(95,53){$2$}
\put(105,55){\line(1,2){18}}
\put(89,15){\circle{6}}
\put(89,15){\line(2,5){16}}
\put(94,13){$4$}
\put(121,15){\circle{6}}
\put(121,15){\line(-2,5){16}}
\put(126,12){$5$}
\put(132,25){\circle{6}}
\put(132,25){\line(2,5){16}}
\put(137,22){$6$}
\put(167,25){\circle{6}}
\put(167,25){\line(-2,5){16}}
\put(170,28){$7$}
\end{picture}
  \caption{An Example of Acyclic graph}\label{fig:S.1}
\end{figure}

Note that, a general walk from node $s$ to node $i$, $s\mapsto i$, can pass through node $i$ multiple times. To express clearly whether a walk passes through node $i$ or not, we denote by $s\stackrel{\backslash i}{\mapsto} i$ a walk from node $s$ to node $i$ without passing through node $i$ in the middle. In particular, $i\stackrel{\backslash i}{\mapsto} i$ a single-return walk for node $i$ (i.e., it goes from node $i$ to node $i$ without node $i$ in the middle). Denote by $(i\stackrel{\backslash i}{\mapsto} i)^{\ell}$ a $\ell$-return walk for node $i$, i.e., it passes through node $i$ for $(\ell-1)$ times in the middle.  

\begin{lem}\label{lem:walk0}
For any node $i\in \mathcal{V}$, $j\in \mathcal{N}_i$ and $k\ge0$, we have
\begin{align}
a_{i\rightarrow j}(k) = \frac{1}{\phi(\{i\mapsto i\} \mid \mathcal{G}_{i\backslash j}(k))}>0 \label{eq:aij}
\end{align}
In addition, $a_{i\rightarrow j}(k) =a_{i\rightarrow j}(d-1)$ for all $k\ge d$, where $d$ is the diameter of $\mathcal{G}$. 
\end{lem}

\begin{IEEEproof}
We first consider a connected graph $\mathcal{G}$ with diameter $d$. We proceed by induction. Note that $\mathcal{G}_{i\backslash j}(k)$ is a depth-$k$ tree graph rooted at node $i$. It is clear that $a_{i\rightarrow j}(0)=1=1/\phi(\{i\mapsto i\} \mid \mathcal{G}_{i\backslash j}(0))>0$ because $\mathcal{G}_{i\backslash j}(0)$ contains $i$ only,  so the only return walk is $(i)$, i.e., $\phi((i))=1$. Now assume, for some $k\ge0$, 
\begin{align*}
a_{i\rightarrow j}(k_1) &= \frac{1}{\phi(\{i\mapsto i\} \mid \mathcal{G}_{i\backslash j}(k_1))}>0
\end{align*}
 for all $i\in \mathcal{V}$, $j\in \mathcal{N}_i$ and all $k_1=0, 1, \ldots, k-1$. We need to show (\ref{eq:aij}) also holds.  To see this, we consider $\mathcal{G}_{i\backslash j}(k)$ and construct its subgraphs $\mathcal{G}_{v\backslash i}(k-1)$ for all $v\in \mathcal{N}_i\backslash j$. Obviously, each such subgraph is depth $k-1$ at most. 
 
Using the notations of $i\stackrel{\backslash i}{\mapsto} i$ and $(i\stackrel{\backslash i}{\mapsto} i)^{\ell}$,  we have 
\begin{align*} 
\phi(\{i\mapsto i\}\mid \mathcal{G}_{i\backslash j}(k)) &=\sum_{\ell} \phi(\{(i\stackrel{\backslash i}{\mapsto} i)^{\ell}\}\mid \mathcal{G}_{i\backslash j}(k))\\
&=\sum_{\ell} \phi(\{i\stackrel{\backslash i}{\mapsto} i\}\mid \mathcal{G}_{i\backslash j}(k))^{\ell}\\
&=\frac{1}{1-\phi(\{i\stackrel{\backslash i}{\mapsto} i\}\mid \mathcal{G}_{i\backslash j}(k))}>0.
\end{align*}
The last equality above and the positivity of the sum follows from the walk-summability assumption. Now, every single-return walk $w$ in $\mathcal{G}_{i\backslash j}(k)$ must be of the form $w=(i, w_v, i)$, where $w_v$ is a return walk (not necessarily single-return) in $\mathcal{G}_{v\backslash i}(k-1)$ for some $v\in \mathcal{N}_i\backslash j$. That is, $\phi(w)=(-a_{iv})(-a_{vi})\phi(w_v)$. Hence, 
\begin{align*}
\phi(\{i\stackrel{\backslash i}{\mapsto} i\}\mid \mathcal{G}_{i\backslash j}(k))
&=\hspace{-1mm}\sum_{v\in \mathcal{N}_i\backslash j} \hspace{-1mm}a_{iv}a_{vi} \phi(\{v\mapsto v\}\mid \mathcal{G}_{v\backslash i}(k-1)) \\
&=\sum_{v\in \mathcal{N}_i\backslash j} \frac{a_{iv}a_{vi}}{a_{v\rightarrow i}(k-1)}.
\end{align*} 
The last step above used the assumption in the induction that $
a_{v\rightarrow i}(k-1) = 1/\phi(\{v\mapsto v\} \mid \mathcal{G}_{v\backslash i}(k-1))$.
Combining the above analysis and using (\ref{ak0}) and (\ref{ak}), we obtain
\begin{align*} 
0&<\frac{1}{\phi(\{i\mapsto i\}\mid \mathcal{G}_{i\backslash j}(k))} \\
&=1-\sum_{v\in \mathcal{N}_i\backslash j} \frac{a_{iv}a_{vi}}{a_{v\rightarrow i}(k-1)} \\&= a_{i\rightarrow j}(k),
\end{align*}
which is (\ref{eq:aij}). The induction proof is complete. The property of $a_{i\rightarrow j}(k) =a_{i\rightarrow j}(d-1)$ for all $k\ge d$ follows from the fact that $\mathcal{G}_{i\backslash j}(k-1)$ has depth $d-1$ at most because $\mathcal{G}$ has diameter of $d$ and $j$ is a neighbour of $i$.  

For the case that $\mathcal{G}$ is not connected, it is composed of a finite number of connected subgraphs, each having a diameter smaller than $d$. It is clear that (\ref{eq:aij}) holds for each subgraph, thus the result in the lemma still holds.  
\end{IEEEproof}

\begin{lem}\label{lem:walk01}
For any node $i\in \mathcal{V}$, $j\in \mathcal{N}_i$ and $k\ge0$, we have
\begin{align}
b_{i\rightarrow j}(k) = \sum_{s}\phi(\{s\stackrel{\backslash i}{\mapsto} i\} \mid \mathcal{G}_{i\backslash j}(k))b_s\label{eq:bij}
\end{align}
where $s$ ranges over all nodes in $\mathcal{G}_{i\backslash j}(k)$.
In addition, $b_{i\rightarrow j}(k) =b_{i\rightarrow j}(d-1)$ for all $k\ge d$, where $d$ is the diameter of $\mathcal{G}$. 
\end{lem}

\begin{IEEEproof}
As in the previous result, it is sufficient to consider a connected $\mathcal{G}$. We proceed by induction. Since $\mathcal{G}_{i\backslash j}(0)$ contains node $i$ only, $\phi(\{s\stackrel{\backslash i}{\mapsto} i\} \mid \mathcal{G}_{i\backslash j}(0))=1$. This means that $\sum_{s}\phi(\{s\stackrel{\backslash i}{\mapsto} i\} \mid \mathcal{G}_{i\backslash j}(0))b_s= b_i=b_{i\rightarrow j}(0)$. That is, (\ref{eq:bij}) holds for $k=0$. 
Now assume that, for some $k\ge1$,  $b_{i\rightarrow j}(k_1) = \sum_{s}\phi(\{s\stackrel{\backslash i}{\mapsto} i\} \mid \mathcal{G}_{i\backslash j}(k_1))b_s$ for all node $i\in \mathcal{V}$, $j\in \mathcal{N}_i$ and $k_1=0, 1, \ldots, k-1$. We need to show (\ref{eq:bij}) holds. Indeed, take any walk $s\stackrel{\backslash i}{\mapsto} i$ in $\mathcal{G}_{i\backslash j}(k)$.  If $s=i$, this walk is just a singleton $(i)$, for which $\phi(\{s\stackrel{\backslash i}{\mapsto} i\} \mid \mathcal{G}_{i\backslash j}(k))b_s=b_i$. For $s\ne i$, this walk must return to $i$ through some $v\in \mathcal{N}_i\backslash j$, i.e., this walk can be written as the concatenation of a sub-walk $s\mapsto v$ in the subgraph $\mathcal{G}_{v\backslash i}(k-1)$ and $(v,\ i)$ for some $v\in \mathcal{N}_i\backslash j$. The subgraph has depth $k-1$ at most because $\mathcal{G}_{i\backslash j}(k)$ has depth $k$ at most. It follows that 
\begin{align*}
&\sum_{s}\phi(\{s\stackrel{\backslash i}{\mapsto} i\} \mid \mathcal{G}_{i\backslash j}(k))b_s \\=&
b_i+\sum_{v\in \mathcal{N}_i\backslash j}\sum_{s} \phi(\{s\mapsto v\} \mid \mathcal{G}_{v\backslash i}(k-1))(-a_{iv})b_s.
\end{align*}
In the right-hand side above, $s$ obviously ranges over all the nodes in $\mathcal{G}_{v\backslash i}(k-1)$. 

Now,  every walk $s\mapsto v$ in $\mathcal{G}_{v\backslash i}(k-1)$ is a concatenation of subwalks $s\stackrel{\backslash v}{\mapsto} v$ and $v\mapsto v$ in $\mathcal{G}_{v\backslash i}(k-1)$. It follows that 
\begin{align*}
&\phi(\{s\mapsto v\} \mid \mathcal{G}_{v\backslash i}(k-1))\\
=& \phi(\{s\stackrel{\backslash v}{\mapsto} v\} \mid \mathcal{G}_{v\backslash i}(k-1))\phi(\{v\mapsto v\} \mid \mathcal{G}_{v\backslash i}(k-1)) \\
=& \phi(\{s\stackrel{\backslash v}{\mapsto} v\} \mid \mathcal{G}_{v\backslash i}(k-1))\frac{1}{a_{v\rightarrow i}(k-1)}. 
\end{align*}
The last step above used Lemma~\ref{lem:walk01}. Combining the analysis above, we get 
\begin{align*}
&\sum_{s}\phi(\{s\stackrel{\backslash i}{\mapsto} i\} \mid \mathcal{G}_{i\backslash j}(k))b_s \\=&
b_i-\sum_{v\in \mathcal{N}_i\backslash j}\sum_{s}\frac{a_{iv}}{a_{v\rightarrow i}(k-1)}\phi(\{s\stackrel{\backslash v}{\mapsto} v\} \mid \mathcal{G}_{v\backslash i}(k-1))b_s\\
=&b_i- \sum_{v\in \mathcal{N}_i\backslash j} \frac{a_{iv}b_{v\rightarrow i}(k-1)}{a_{v\rightarrow i}(k-1)}\\
=&b_{i\rightarrow j}(k).
\end{align*}
The second last step above used the induction assumption, and the last step used (\ref{bk0}) and (\ref{bk}). This completes the induction proof. Finally, $b_{i\rightarrow j}(k) =b_{i\rightarrow j}(d-1)$ for all $k\ge d$ follows from the fact that $\mathcal{G}_{i\backslash j}(k)$ has depth $d-1$ at most. 
\end{IEEEproof}

Now we are ready to prove Theorem~\ref{thm:1}. 
\begin{IEEEproof}
As before, it suffices to consider the case that $\mathcal{G}$ is connected. The first two properties in the theorem follow directly from Lemmas~\ref{lem:walk0}-\ref{lem:walk01}. It remains to show (\ref{xid}). 

\begin{figure}[ht]
\begin{picture}(240,95)
\put(124,93){\circle{6}}
\put(124,93){\line(1,-1){25}} 
\put(150,68){\circle{6}}
\put(155,66){$3$}
\put(129,91){$1$}
\put(105,55){\circle{6}}
\put(95,53){$2$}
\put(105,55){\line(1,2){18}}
\put(89,15){\circle{6}}
\put(89,15){\line(2,5){16}}
\put(94,13){$4$}
\put(121,15){\circle{6}}
\put(121,15){\line(-2,5){16}}
\put(126,12){$5$}
\put(132,25){\circle{6}}
\put(132,25){\line(2,5){16}}
\put(137,22){$6$}
\put(167,25){\circle{6}}
\put(167,25){\line(-2,5){16}}
\put(170,28){$7$}
\put(152,71){\line(1,1){20}}
\put(173,92){\circle{6}}
\put(179,91){$0$}
\end{picture}
  \caption{Acyclic graph for with a fictitious node 0}\label{fig:S.1.0}
  
\end{figure}
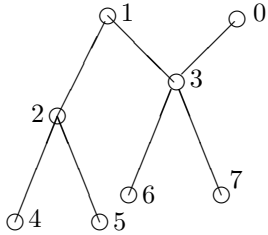
Consider any node $i\in \mathcal{V}$. Add a fictitious node 0 and fictitious edge $(i, \ 0)$ to the graph $\mathcal{G}$ and denote the augmented graph by $\tilde{\mathcal{G}}$, as depicted in Fig.~\ref{fig:S.1.0} for root node $i=3$. Let $k\ge d_i$, where $d_i$ is the depth of the graph $\mathcal{G}$ when treating node $i$ as the root. It is clear that $d_i\le d$ and that $\tilde{\mathcal{G}}_{i\backslash 0}(k) = \mathcal{G}$. We have from (\ref{ak0}) and (\ref{ak}), 
\begin{align*}
a_{i\rightarrow 0}(k) = a_{ii} - \sum_{v\in \mathcal{N}_i} \frac{a_{vi}a_{iv}}{a_{v\rightarrow i}(k-1)} = \tilde{a}_i(k).
\end{align*}
On the other hand, from Lemma~\ref{lem:walk0}, we have
\begin{align*}
a_{i\rightarrow 0}(k) &= \frac{1}{\phi(\{i\rightarrow i\} \mid \tilde{\mathcal{G}}_{i\backslash 0}(k))} \\
&=\frac{1}{\phi(\{i\rightarrow i\} \mid \mathcal{G})} \\
&= \frac{1}{\phi(\{i\rightarrow i\})}.
\end{align*}
It follows that 
\begin{align*}
\tilde{a}_i(k)=\frac{1}{\phi(\{i\rightarrow i\})}.
\end{align*}
Similarly, from (\ref{bk0}) and (\ref{bk}), 
\begin{align*}
b_{i\rightarrow 0}(k) = b_i - \sum_{v\in \mathcal{N}_i} \frac{a_{iv}b_{v\rightarrow i}(k-1)}{a_{v\rightarrow i}(k-1)} = \tilde{b}_i(k).
\end{align*}
On the other hand, from Lemma~\ref{lem:walk01}, we have
\begin{align*}
b_{i\rightarrow 0}(k)&= \sum_s \phi(\{s\stackrel{\backslash i}{\mapsto} i\} \mid \tilde{\mathcal{G}}_{i\backslash 0}(k))b_s \\ &=\sum_s \phi(\{s\stackrel{\backslash i}{\mapsto} i\} \mid \mathcal{G})b_s\\
&=\sum_s \phi(\{s\stackrel{\backslash i}{\mapsto} i\})b_s.
\end{align*}
It follows that 
\begin{align*}
\tilde{b}_i(k)=\sum_s \phi(\{s\stackrel{\backslash i}{\mapsto} i\})b_s.
\end{align*}
Also note $\phi(\{s\mapsto i\})=\phi(\{s\stackrel{\backslash i}{\mapsto} i \})\phi(\{i\mapsto i\})$.

Finally, combining the above with Lemma~\ref{lem:key}, we get
\begin{align*}
x_i^{\star} &= \sum_s \phi(\{s\mapsto i\}) b_s  \\
&= \sum_s \phi(\{s\stackrel{\backslash i}{\mapsto} i \})\phi(\{i\mapsto i\})  b_s  \\
&=  \sum_s \frac{1}{\tilde{a}_i(k)}\phi(\{s\stackrel{\backslash i}{\mapsto} i \})  b_s  \\
&= \frac{\tilde{b}_i(k)}{\tilde{a}_i(k)} = \hat{x}_i(k).
\end{align*}
The last step above comes from (\ref{xik}). Since the above holds for all $k\ge d_i$, hence for all $k\ge d$, this completes the proof. 
\end{IEEEproof}

\begin{rem}\label{rem:10}
We see in the proof above that $\tilde{a}_i(k)$ and $\tilde{b}_i(k)$ actually converge after $d_i$ iterations, where $d_i$ is the depth of the graph $\mathcal{G}$ when treating node $i$ as the root. This property will be useful in the proof of Theorem~\ref{thm:2}.   
\end{rem}

\section{Convergence Analysis for Loopy Graphs}\label{sec:loopy}

In this section, we consider a general graph $\mathcal{G}$ (acyclic or loopy) under the walk-summability assumption to prove Theorem~\ref{thm:2}. Similar to the proof of Theorem~\ref{thm:1}, it suffices to consider a connected graph $\mathcal{G}$, which we will assume below. We will also assume that $a_{ii}=1$ for all $i$, as the case of $a_{ii}\ne 1$ does not alter the proof, but just complicates the writing.

\subsection{Unwrapped Graph}

Following the work of~\cite{Weiss}, we construct an {\em unwrapped tree} with {\em depth} $t>0$ for a loopy graph $\mathcal{G}$~\cite{Weiss}. Take node $i$ to be the root and then iterate the following procedure $t$ times:
\begin{itemize}
\item Find all leaves of the tree (start with the root);
\item For each leaf, find all the nodes in the loopy graph that neighbor this leaf node, except its parent node in the tree, and add all these node as the children to this leaf node.
\end{itemize}

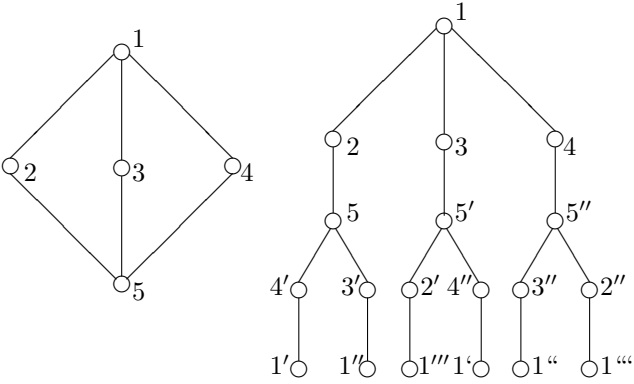
\begin{figure}[ht]
\begin{picture}(240,135)
\put(54,124){\circle{6}}
\put(57,123){\line(1,-1){39}} 
\put(51,123){\line(-1,-1){39}} 
\put(54,121){\line(0,-1){38}} 
\put(54,80){\circle{6}}
\put(12,81){\circle{6}}
\put(96,81){\circle{6}}
\put(54,77){\line(0,-1){38}}
\put(56,38){\line(1,1){40}} 
\put(52,38){\line(-1,1){40}} 
\put(54,36){\circle{6}}
\put(58,30){$5$}
\put(58,126){$1$}
\put(58,75){$3$}
\put(17,75){$2$}
\put(99,75){$4$}

\put(176,134){\circle{6}}
\put(179,133){\line(1,-1){39}} 
\put(173,133){\line(-1,-1){39}} 
\put(176,131){\line(0,-1){38}} 
\put(176,90){\circle{6}}
\put(134,91){\circle{6}}
\put(218,91){\circle{6}}
\put(176,87){\line(0,-1){25}}
\put(134,88){\line(0,-1){25}}
\put(218,88){\line(0,-1){25}}
\put(134,60){\circle{6}}
\put(176,60){\circle{6}}
\put(218,60){\circle{6}}
\put(180,136){$1$}
\put(180,85){$3$}
\put(139,85){$2$}
\put(221,85){$4$}
\put(139,60){$5$}
\put(180,60){$5'$}
\put(222,60){$5''$}
\put(133,57){\line(-3,-5){12}}
\put(135,57){\line(3,-5){12}}
\put(175,57){\line(-3,-5){12}}
\put(177,57){\line(3,-5){12}}
\put(217,57){\line(-3,-5){12}}
\put(219,57){\line(3,-5){12}}
\put(121,34){\circle{6}}
\put(147,34){\circle{6}}
\put(163,34){\circle{6}}
\put(190,34){\circle{6}}
\put(205,34){\circle{6}}
\put(231,34){\circle{6}}
\put(121,31){\line(0,-1){24}}
\put(147,31){\line(0,-1){24}}
\put(163,31){\line(0,-1){24}}
\put(190,31){\line(0,-1){24}}
\put(205,31){\line(0,-1){24}}
\put(231,31){\line(0,-1){24}}
\put(121,4){\circle{6}}
\put(147,4){\circle{6}}
\put(163,4){\circle{6}}
\put(190,4){\circle{6}}
\put(205,4){\circle{6}}
\put(231,4){\circle{6}}
\put(110,32){$4'$}
\put(137,32){$3'$}
\put(167,32){$2'$}
\put(177,32){$4''$}
\put(209,32){$3''$}
\put(235,32){$2''$}
\put(110,2){$1'$}
\put(136,2){$1''$}
\put(166,2){$1'''$}
\put(179,2){$1`$}
\put(209,2){$1``$}
\put(235,2){$1```$}
\end{picture}
  \caption{Left: A loopy graph. Right: The unwrapped tree for root node 1 with 4 layers ($t=4$)}\label{fig:2}
\end{figure}

The variables and weights for each node in the unwrapped tree are copied from the corresponding nodes in the loopy graph. 
It is clear that taking each node as root node will generate a different unwrapped tree.  Fig.~\ref{fig:2} shows the unwrapped tree around root node 1 for a loopy graph. Note, for example, that nodes $1', 1'',1''', 1`, 1``,1```$ all carry the same values $b_1$ and $a_{11}$. Similarly, if node 1' is the parent (or child) of node $j'$ in the unwrapped tree, and node 1 and node $j$ are a wrapped version of nodes 1 and $j$, then $a_{1'j'}=a_{1j}$ (or $a_{j'1'}=a_{j1}$). 

List the nodes in the unwrapped tree in {\em breadth first} order, by starting from the root node, followed by the first layer (i.e., the children of the root node), then the second layer, etc. Denote the unwrapped tree as $\mathcal{G}_i^{(t)}=\{\mathcal{V}_i^{(t)}, \mathcal{E}_i^{(t)}\}$ with the associated matrix $A_i^{(t)}=I-R_i^{(t)}$.  It is obvious that $\mathcal{G}_i^{(t)}$ is connected by construction.  

We have the following key result from \cite{Mailoutov}. 

\begin{lem}\label{lem:loopy1}
There is a one-to-one correspondence between finite-length walks in $\mathcal{G}$ that end at $i$, and walks in $\mathcal{G}_i^{(\infty)}$.  
\end{lem}

We are now ready to prove Theorem~\ref{thm:2}. 

\begin{IEEEproof}
Consider any $i\in \mathcal{V}$. Recall from (\ref{eq:walk_xi}) that $x_i^{\star} = \sum_s \phi(\{s\mapsto i\}) b_s$. Take any $s\in \mathcal{V}$ and consider the set of walks $\{s\mapsto i\}$. From Lemma~\ref{lem:loopy1}, any finite-length walk $w$ in $\{s\mapsto i\}$ (that starts in $s$) has a counterpart in $\mathcal{G}_i^{(t)}$ for a sufficiently large $t$ and this walk starts with $s$ or a replica of it. With some abuse of notion, we denote by $\mathcal{W}(s\mapsto i)^{(t)}$ the set of walks in $\mathcal{G}_i^{(t)}$ that ends at $i$ but starts from either $s$ or a replica of it. Lemma~\ref{lem:loopy1} also implies that any walk (finite-length or not) in $\mathcal{W}(s\mapsto i)^{(t)}$ (which starts with $s$ or a replica of it) has a counterpart in $\mathcal{G}$ that starts in $s$.  This last property also implies that the set of walks $\mathcal{W}(s\mapsto i)^{(t)}$ is walk-summable. Due to the nesting of $\mathcal{G}_i^{(t)}\subset \mathcal{G}_i^{(t+1)}$, we have $\mathcal{W}(s\mapsto i)^{(t)} \subset \mathcal{W}(s\mapsto i)^{(t+1)}$. Therefore, Lemma~\ref{lem:loopy1}  implies that 
\begin{align}
\phi(\{s\mapsto i\}) &= \phi (\bigcup_t \mathcal{W}(s\mapsto i)^{(t)}) = \lim_{t\rightarrow \infty}  \phi (\mathcal{W}(s\mapsto i)^{(t)}). \label{eq:tempxxx}
\end{align} 
The last step follows from Property P3 of walk-summability. 

On the other hand, by defining
\begin{align*}
x_i^{(t)} &= \sum_{s} \phi (\mathcal{W}(s\mapsto i)^{(t)})b_s,
\end{align*}
Lemma~\ref{lem:key} says that $x_i^{(t)}$ is the correct solution of $x_i$ for the graph $\mathcal{G}_i^{(t)}$. Note that $\mathcal{G}_i(t)$ has depth $t$ when viewing from root node $i$. By applying Theorem~\ref{thm:1} to $\mathcal{G}_i^{(t)}$ and noting Remark~\ref{rem:10}, we know that $x_i^{(t)}$ is correctly obtained after running Algorithm~\ref{alg1} on $\mathcal{G}_i^{(t)}$ for $t$ iterations. Also, by the construction of the unwrapped tree graph, we know that running Algorithm~\ref{alg1} on $\mathcal{G}_i^{(t)}$ for $t$ iterations is the same as running Algorithm~\ref{alg1} on $\mathcal{G}$ for $t$ iterations, and the latter generates $\hat{x}_i(t)$ in (\ref{xik}). Hence,  $x_i^{(t)}=\hat{x}_i(t)$. Combining this with (\ref{eq:tempxxx}), we conclude that
 \begin{align*}
 \lim_{t\rightarrow \infty} \hat{x}_i(t) &=  \lim_{t\rightarrow \infty} \sum_{s} \phi (\mathcal{W}(s\mapsto i)^{(t)})b_s\\
 &=\sum_{s} \phi(\{s\mapsto i\})b_s = x_i^{\star},
 \end{align*}
which is (\ref{loop_xik2}). The last step above used Lemma~\ref{lem:key}. 
\end{IEEEproof}

\section{Examples}\label{simulation}

The convergence of Algorithm~\ref{alg1} is demonstrated in this section through two examples, one on an acyclic graph and one on a loopy graph. The performance is compared with two methods. The first method is the Jacobi method (\ref{Jac}) as explained earlier.  Incidentally, the Jacobi method is guaranteed to converge correctly for walk-summable systems because, using $A=D_A(I-R)$,  (\ref{Jac}) can be rewritten as
\begin{align}
\hat{x}(k+1)=R\hat{x}(k)+D_A^{-1}b,  \label{Jac1}
\end{align} 
and we have $\rho(R)<1$. 

The second method we consider is the distributed algorithm in \cite{Morse} for solving linear systems. This algorithm is generalized from a Laplacian matrix based consensus approach by applying orthogonal projection. Defining $\mathbf{x}_i(k)$ as the (vector) estimate of $x^{\star}$ by node $i$ at iteration $k$, the distributed algorithm in \cite{Morse} deploys the following distributed iteration:
\begin{align}
\mathbf{x}_i(k+1) &=\mathbf{x}_i(k)-\frac{1}{|\mathcal{N}_i|} P_i\left (|\mathcal{N}_i|\mathbf{x}_i(k)-\sum_{j\in\mathcal{N}_i} \mathbf{x}_j(k)\right ) \label{xxxxx}
\end{align}
with the initial condition of $\mathbf{x}_i(0)$ to be any solution for the $i$-th row of (\ref{Axb}), where $P_i$ is an orthogonal projection matrix for the $i$-th row. In particular, we take 
$\mathbf{x}_i(0)=\frac{b_i}{a_{ii}} e_i$ with $e_i$ being the vector with $i$-th component equal to 1 and all other components equal to 0. Note that this algorithm does not satisfy the low complexity constraints C2-C3 because each node $i$ estimates the whole vector of $x$, thus not qualifying for a distributed algorithm in our definition. Nevertheless, this algorithm is used to compare convergence. 

\subsection{Example 1: Acyclic Graph}

Our first example is an acyclic graph depicted in Fig.~\ref{fig:S.1}. The matrix $A$ has $a_{ii}=|\mathcal{N}_i|$ and the non-zero $a_{ij}$ randomly chosen from $(-1, \ -0.85)$ and $b_i=i$. The diameter of the graph is $d=4$. The simulated result for Algorithm~\ref{alg1} is shown in Fig.~\ref{fig:S.2}. As we can see, it takes only $4$ iterations to converge. In comparison, we show in Fig.~\ref{fig:S.3} the simulated result for the same example using the Jacobi method and in Fig.~\ref{fig:S.4} using the distributed method for linear systems in \cite{Morse}. As we see that the Jacobi method converges in about 60 iterations, with an error of approximately 0.1. For reaching a similar error of 0.1, it takes the method in \cite{Morse} about 5000 iterations. 

\begin{figure}
\begin{center}
\includegraphics[width=8.0cm]{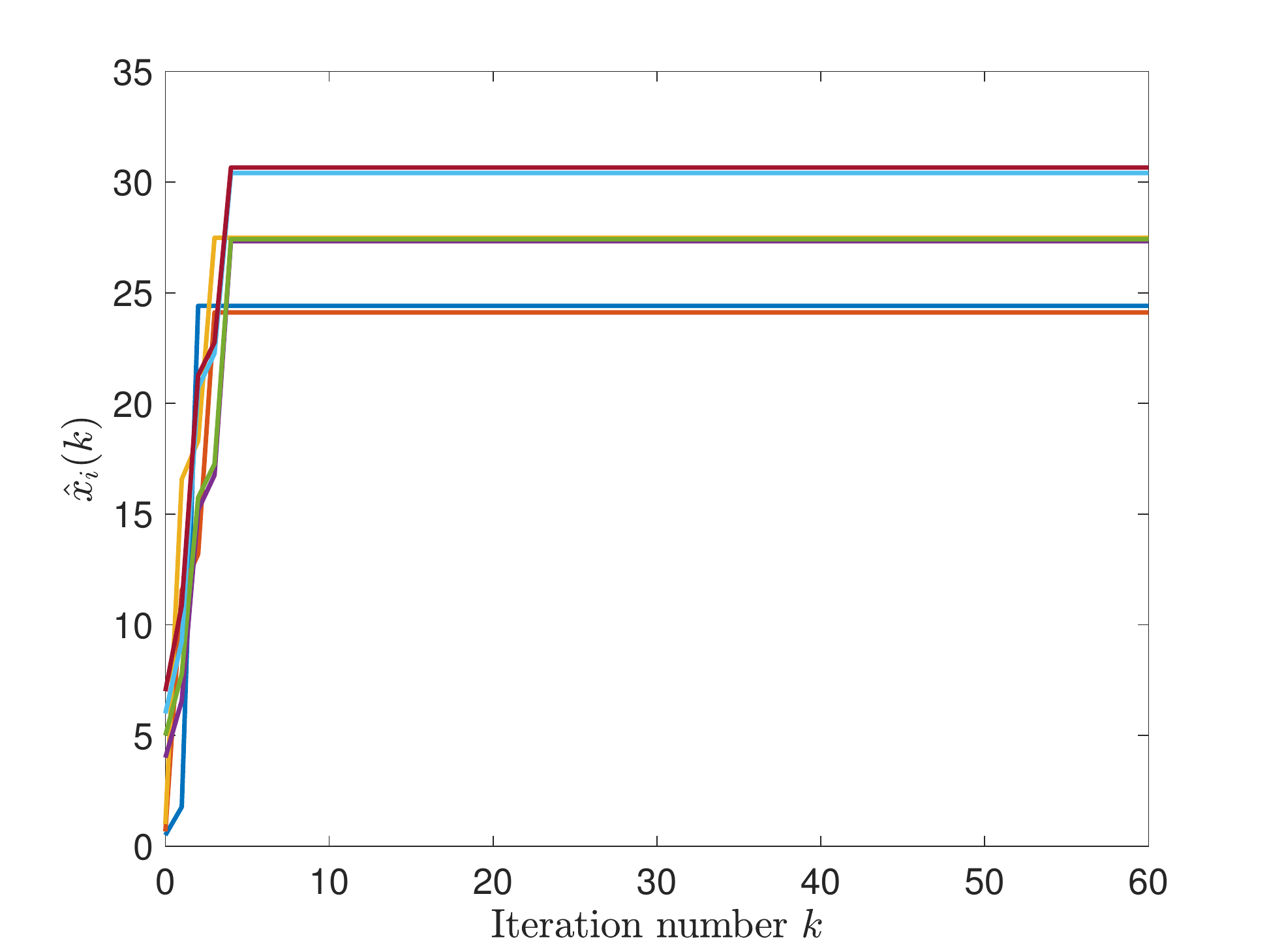}
\end{center}
  \caption{Convergence of the Proposed Algorithm~\ref{alg1} on acyclic graph}\label{fig:S.2}
\end{figure}
\begin{figure}
\begin{center}
  \includegraphics[width=8.0cm]{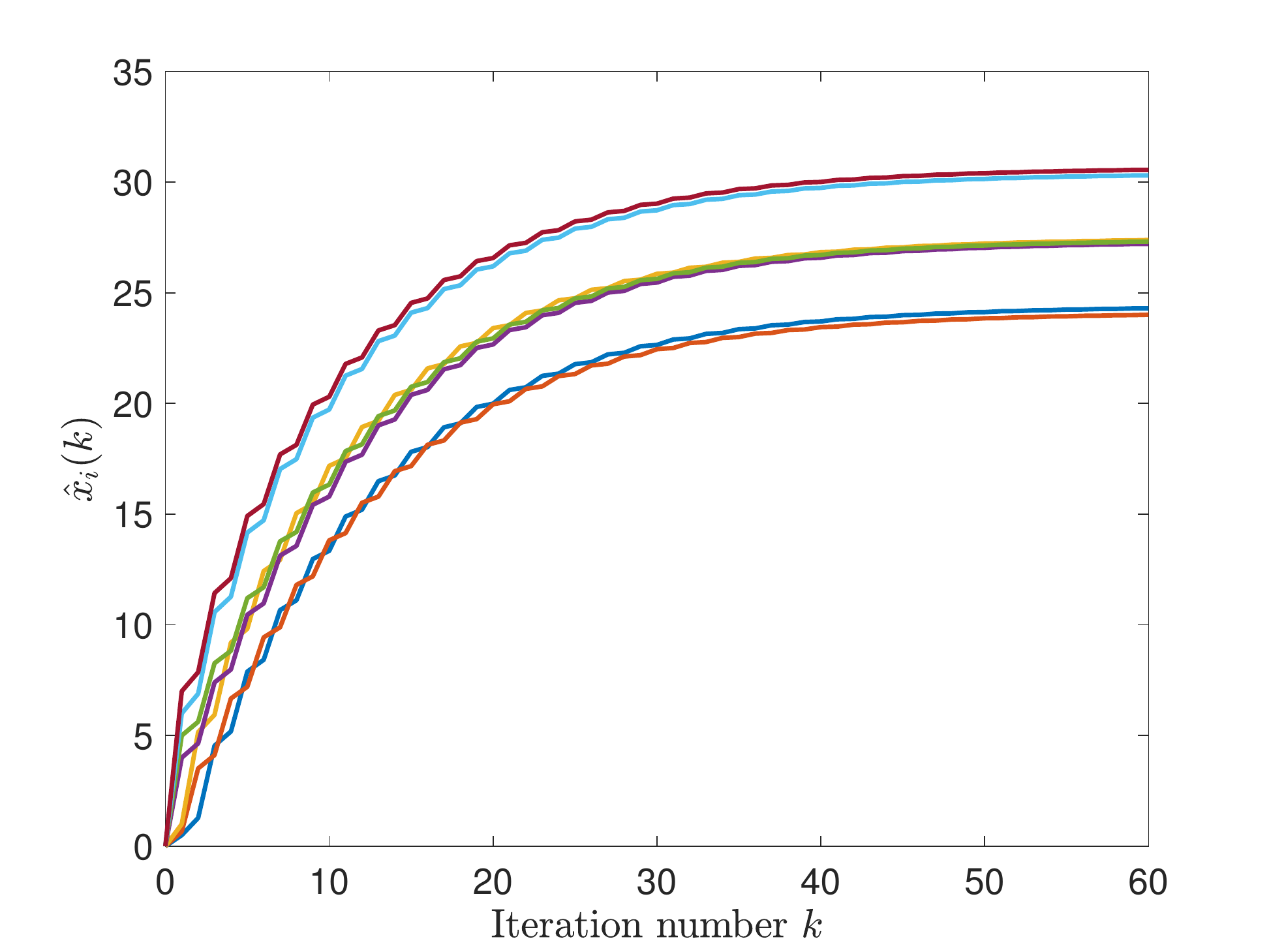}
\end{center}
  \caption{Convergence of Jacobi method on acyclic graph}\label{fig:S.3}
\end{figure}
\begin{figure}
\begin{center}
  \includegraphics[width=8.0cm]{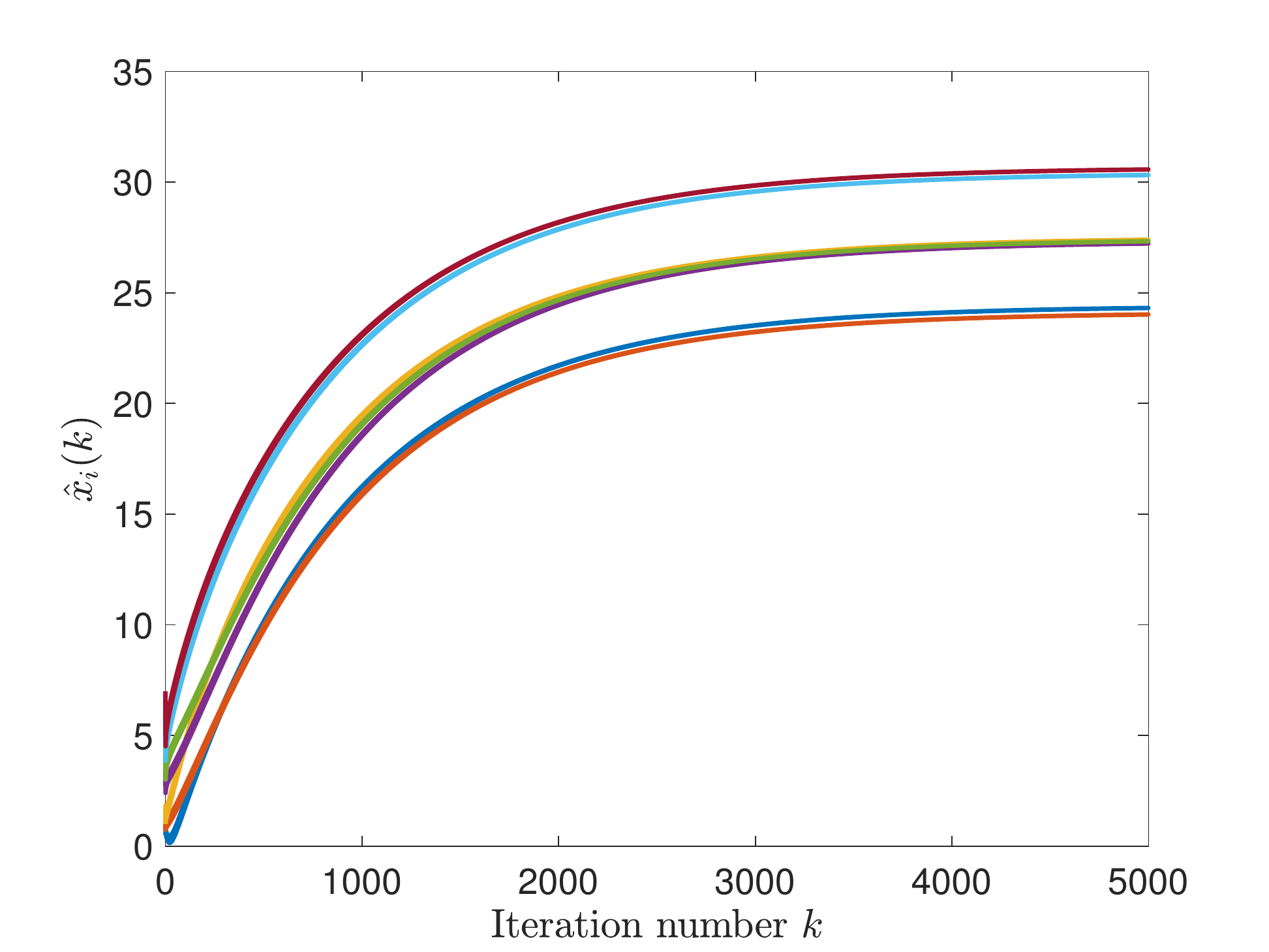}
\end{center}
  \caption{Convergence of the distributed method in \cite{Morse} on acyclic graph}\label{fig:S.4}
\end{figure}

\subsection{Example 2: Loopy Graph}

Our second example is a loopy graph depicted in Fig.~\ref{fig:S.5}. The values of $A$ and $b_i=i$ are chosen in the same way as before. This choice makes $A$ diagonally dominant, hence walk-summable. The simulated result for Algorithm~\ref{alg1} is shown in Fig.~\ref{fig:S.6}. For 100 iterations, the error is converged down to approximately $0.8\times10^{-4}$. Fig.~\ref{fig:S.7} shows the simulated result for the Jacobi method and Fig.~\ref{fig:S.8} is the distributed method in \cite{Morse}. As we see that the Jacobi method converges considerably slower, with an error of approximately 0.02 after 100 iterations. For reaching a similar error, it takes the method in \cite{Morse} nearly 20,000 iterations. 

\begin{figure}
\begin{center}
  \includegraphics[width=6cm]{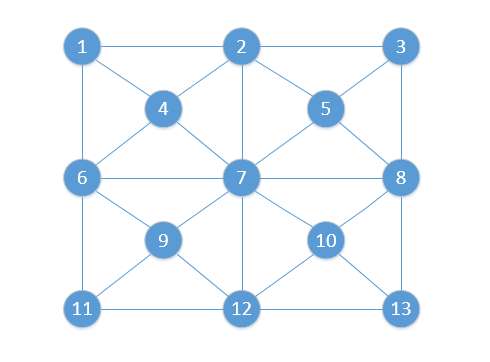}
\end{center}
  \caption{A 13-node Loopy Graph for Example 2}\label{fig:S.5}
\end{figure}

\begin{figure}
\begin{center}
  \includegraphics[width=8.0cm]{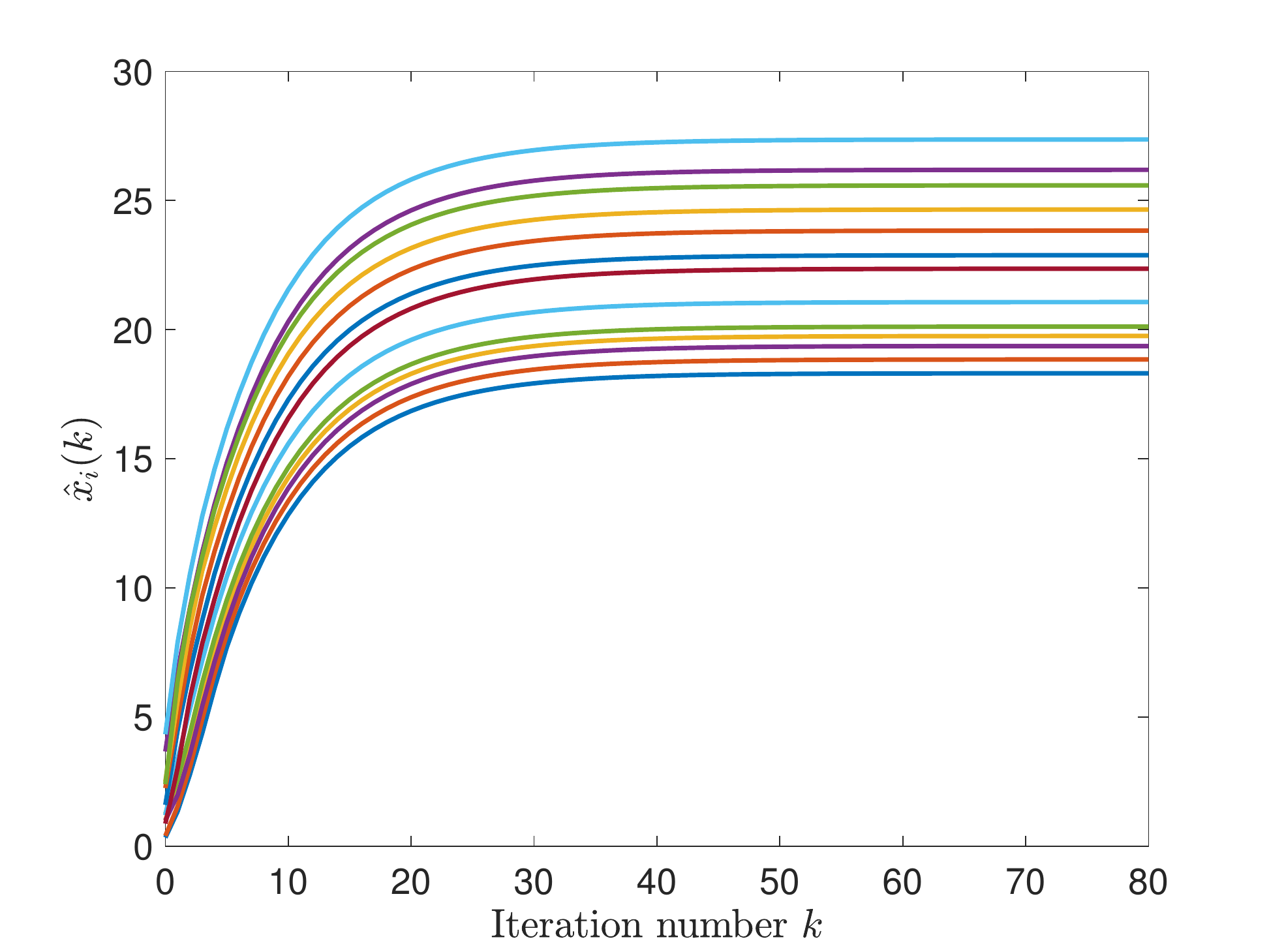}
\end{center}
  \caption{Convergence of the Proposed Algorithm~\ref{alg1} on loopy graph}\label{fig:S.6}
\end{figure}
\begin{figure}
\begin{center}
  \includegraphics[width=8.0cm]{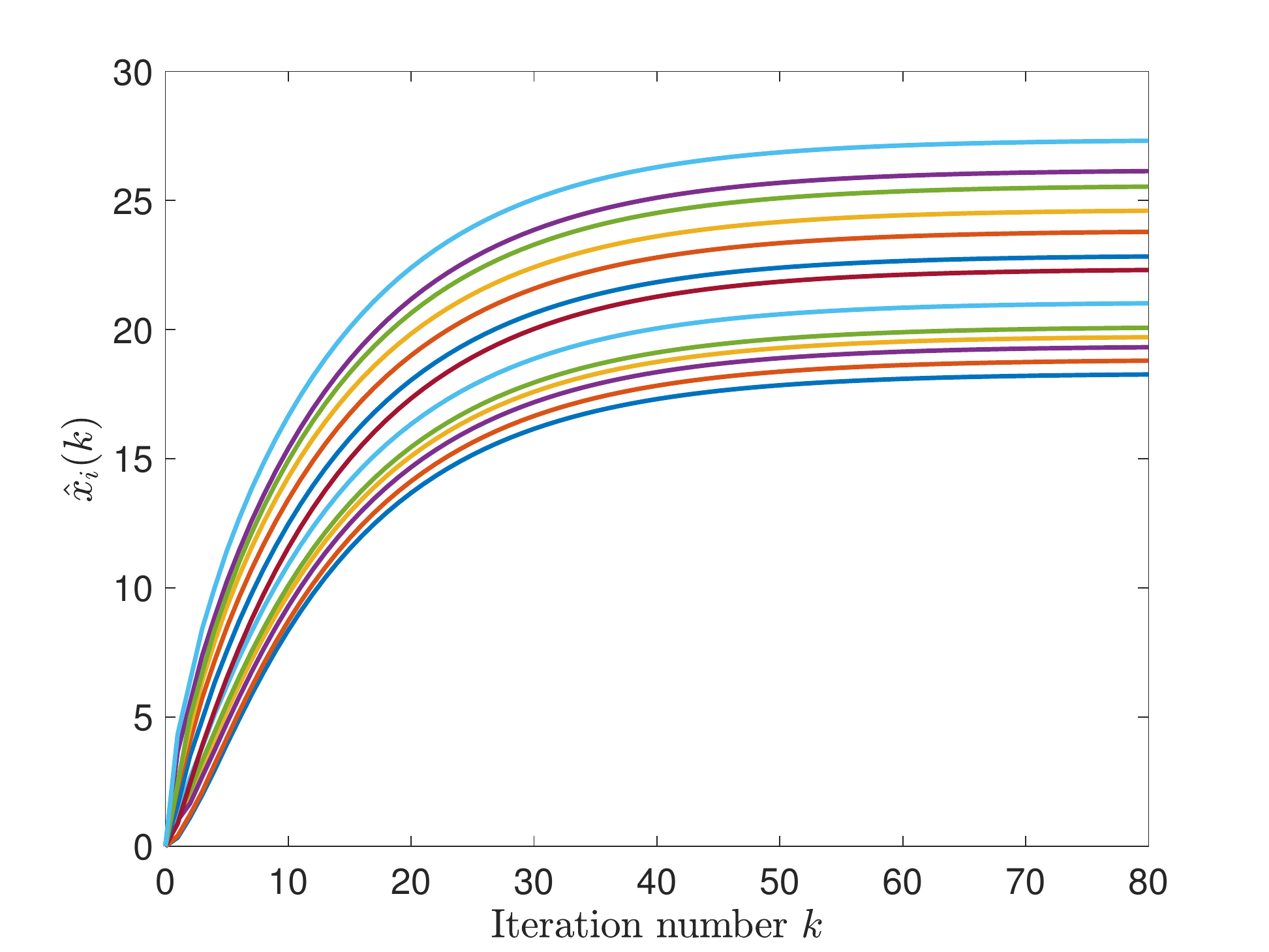}
\end{center}
  \caption{Convergence of the Jacobi method on loopy graph}\label{fig:S.7}
\end{figure}
\begin{figure}
\begin{center}
  \includegraphics[width=8.0cm]{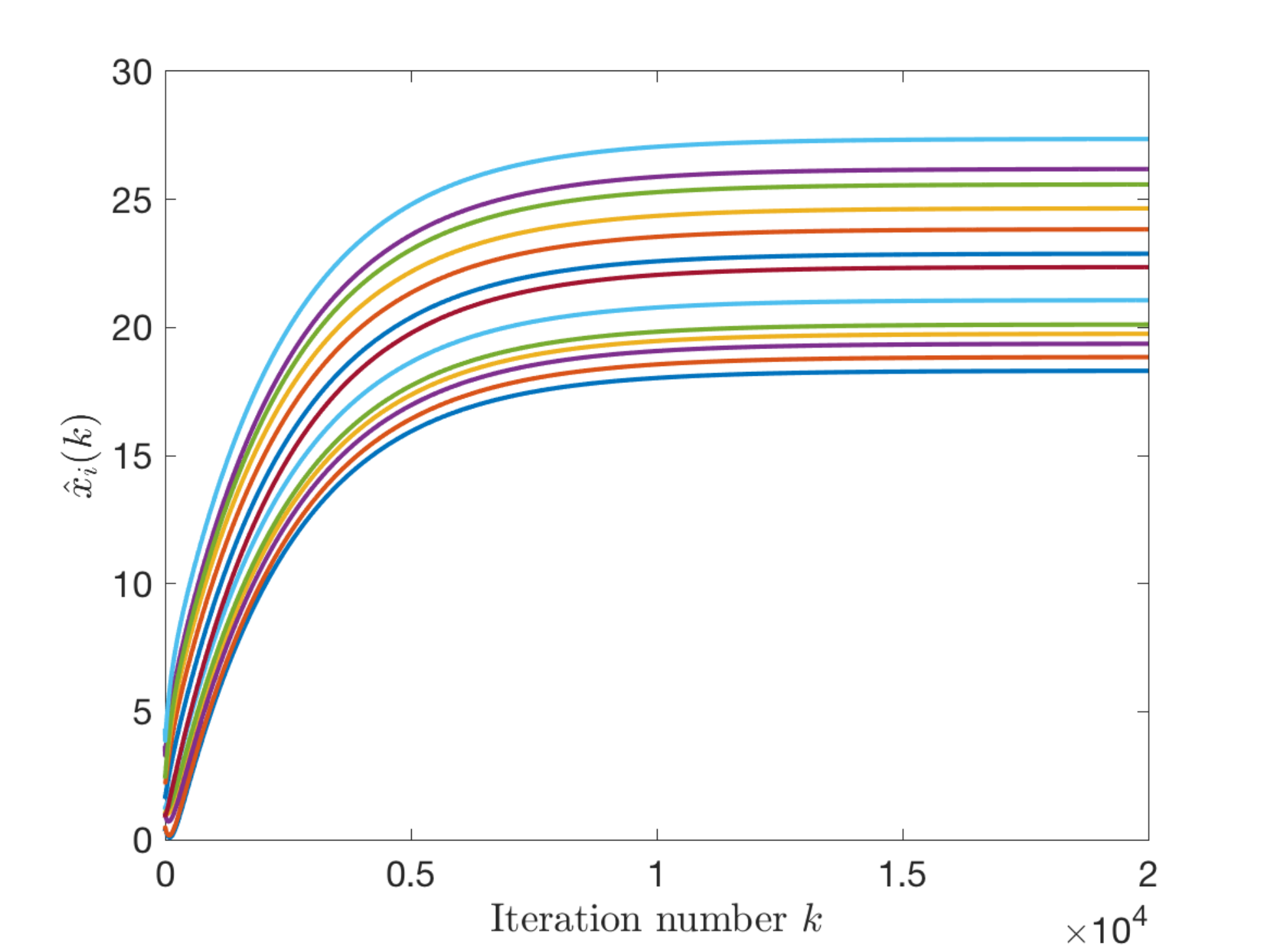}
\end{center}
  \caption{Convergence of the distributed method in \cite{Morse} on loopy graph}\label{fig:S.8}
\end{figure}

\subsection{Example 3: Large Loopy Graph}
Our third example involves a randomly connected graph with 1000 nodes, as shown in Fig.~\ref{fig:S.10}. The circles indicate the nodes and the curves indicated the edges. The matrix $A$ is chosen by taking the non-zero off-diagonal terms $R_{ij}$ (the $(i,j)$-th entry of $R$) to be random within $(-0.05, \ \ 0.05)$, and $b_i=i$ for all $i$.  Fig.~\ref{fig:S.11-1} plots the convergence of the $\hat{x}(k)$. To better show the convergence of Algorithm~\ref{alg1}, we plot the logarithmic error $\log_{10}(\|\hat{x}(k) - x^{\star}\|^2/n)$ versus the iteration number $k$ in Fig.~\ref{fig:S.11}. As we can see from the figure, the error decreases exponentially (or the logarithmic error decreases linearly) in the number of iterations. The error floor after 15 iterations or so is due to the accuracy limitation of Matlab. Measuring from the figure, the decreasing slope is roughly equal to -1. In comparison, for this  specific realisation of $R$, the computation shows that $\log_{10}(\rho(R))=-0.9395$ and $\log_{10}(\rho(\bar{R})) =-0.5167$.
 
\begin{figure}
\begin{center}
  \includegraphics[width=8.0cm]{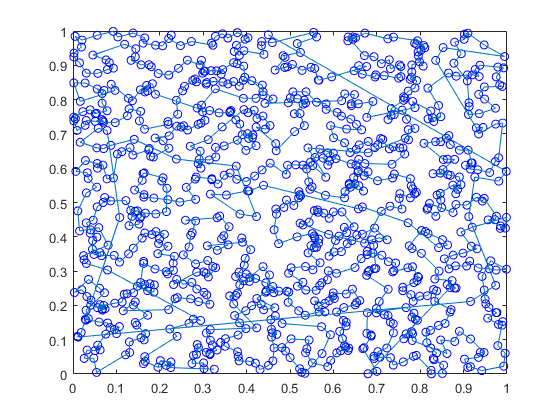}
\end{center}
  \caption{A 1000-node Loopy Graph for Example 3}\label{fig:S.10}
\end{figure}

\begin{figure}
\begin{center}
  \includegraphics[width=8.0cm]{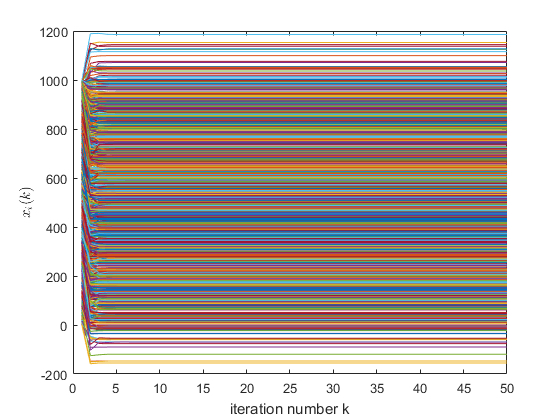}
\end{center}
  \caption{Convergence of the Proposed Algorithm~\ref{alg1}}\label{fig:S.11-1}
\end{figure}

\begin{figure}
\begin{center}
  \includegraphics[width=8.0cm]{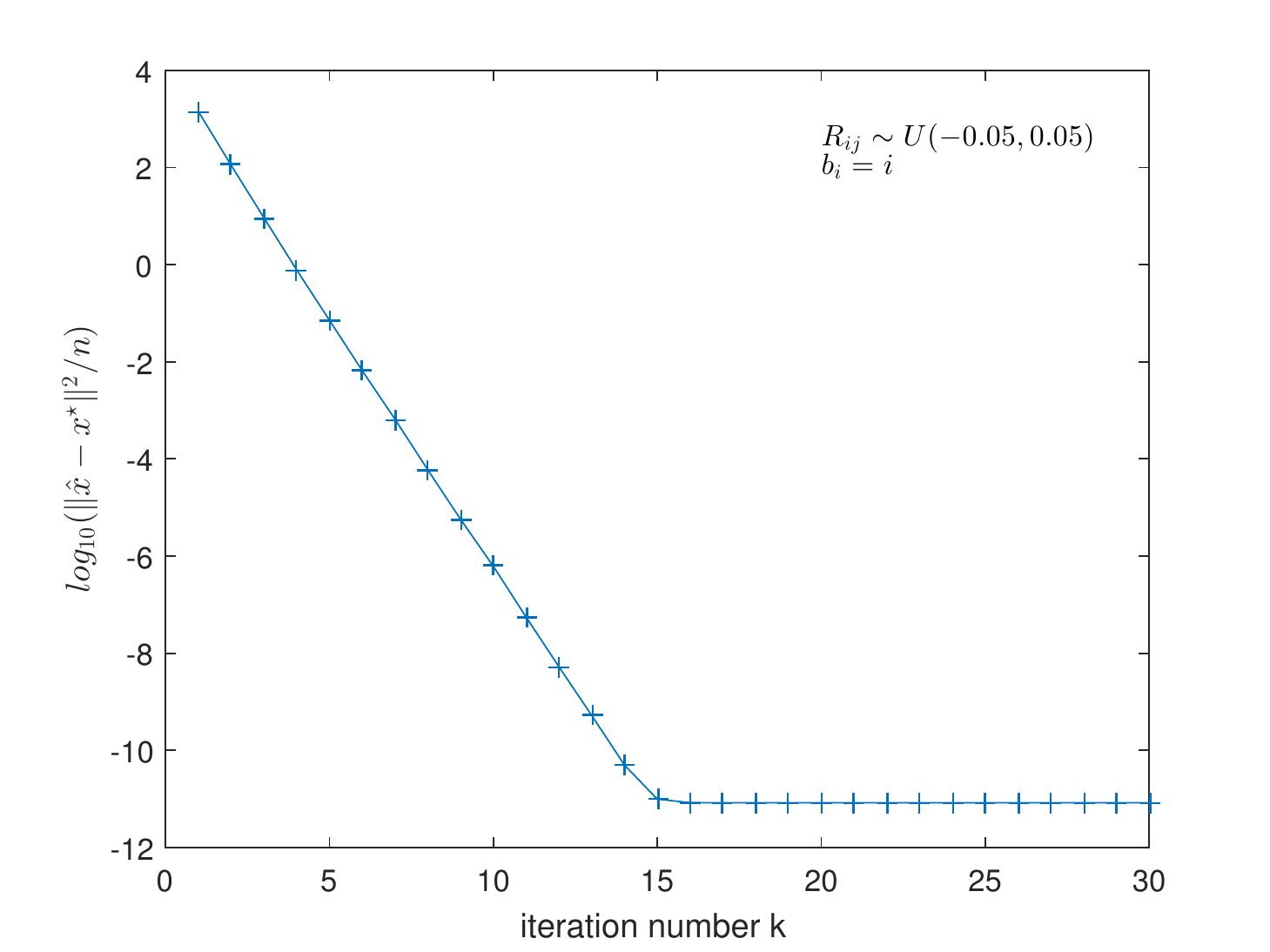}
\end{center}
  \caption{Convergence Error of the Proposed Algorithm~\ref{alg1}}\label{fig:S.11}
\end{figure}

\begin{rem}\label{rem:BP-Jac}
To help understand why Algorithm~\ref{alg1} outperforms the Jacobi method, it is straightforward to derive from (\ref{Jac1}) that 
\begin{align*}
\hat{x}(k)&=\sum_{\ell =1}^k R^{\ell} (D_A^{-1}b).
\end{align*}
The above means that in the $k$-th iteration, the Jacobi estimate $\hat{x}_i(k)$ in (\ref{Jac1}) will contain all the walks (from every node $s$) to node $i$ up to length $k$. In other words, the Jacobi estimate $\hat{x}_i(k)$ in (\ref{Jac1}) misses all the walks of lengths $k+1$ or longer, when compared with its asymptotic value (i.e., the exact value of $x_i^{\star}$). In comparison, a closer look of the analysis in Section~\ref{sec:loopy} can reveal that the proposed estimate $\hat{x}_i(k)$ in Algorithm~\ref{alg1} actually contains {\em all} the walks in the unwrapped depth-$k$ tree $\mathcal{G}_i^{(k)}$, rooted at node $i$, including all the walks used by the Jacobi method as a strict subset.  This explains why Algorithm~\ref{alg1} outperforms the Jacobi method in general.
\end{rem}

\section{Conclusion}
We take this space to comment on several theoretical features of the proposed distributed algorithm for linear systems. 

Firstly, the algorithm can be viewed as an enhanced version of the classical Jacobi method. In the Jacobi method, the information flow from node $i$ to its neighboring node $j$ in iteration $k$ gets fed back right away in iteration $k+1$. That is, $\hat{x}_i(k)$ is used in computing $\hat{x}_j(k+1)$, which is then used in computing $\hat{x}_i(k+2)$, as evident in (\ref{Jac}).  Such direct feedback creates information contamination, and it requires an infinite number of iterations for the contamination to dissipate completely. In the proposed algorithm, the information flow from node $i$ to node $j$ moves on to other nodes without returning to node $i$ directly. This allows the algorithm to converge only in a finite (and minimum) number of iterations for an acyclic graph. Even for a cyclic graph, this feature leads to a much faster convergence because the information can flow back only through loops, resulting in much weaker information contamination. 

The tradeoff is that a slightly stronger condition, i.e., walk-summability (or $\rho(\bar{R})<1$) is required to guarantee the correct convergence of the proposed algorithm, whereas the Jacobi method requires only $\rho(R)<1$. The technical insight for this difference is that unordered sums of walks are used in the proposed algorithm, whereas the Jacobi method uses a specific ordered sums of walks, with the ordering given by (\ref{Jac}). 

Secondly,  the proposed algorithm can be viewed as a distributed version of Gauss elimination method. The link to Gauss elimination is evident from the proofs of Theorems~\ref{thm:1}-\ref{thm:2}. Effectively, every node carries out its own sequence of Gauss eliminations independently by moving required information around through nodes in the graph. 

Thirdly, the proposed algorithm resembles the well-known Belief Propagation (BP) algorithm for computing the marginal distributions of a multi-variate probability density function, originally proposed by Pearl~\cite{Pearl}. Indeed, some ideas are adopted from the BP algorithm to develop the proposed algorithm and its convergence proof for the loopy graph case is partially based on an idea in the important works of \cite{Weiss} and \cite{Mailoutov}, namely, the use of unwrapped tree graphs and walk-summability analysis. 

Finally, it would be interesting and important to generalise the proposed algorithm to block diagonally matrices, i.e., each $x_i$ is a vector rather than a scalar. Proper development of this generalization will be carried out in the future.  Finding weaker conditions than walk summability for correct convergence will be an important task in the future. Also, finding out the exact convergence rate of the proposed algorithm will be very interesting too, and we conjecture that the convergence rate is theoretically faster than that of the Jacobi method. Generalising our work to the scenario when the system data $(A,b)$ and graph topology $\mathcal{G}$ are time-varying  will be very interesting for many real-time applications as well. 

\section*{Appendix: Walk Summability Properties}

We have the following result on walk summability, generalised from \cite{Mailoutov} for symmetric and positive  semidefinite matrices to general matrices in our case. 
\begin{lem}\label{lem:walk}
Given an $n\times n$ matrix $R=\{r_{ij}\}$ and its induced graph $\mathcal{G}=\{\mathcal{V}, \mathcal{E}\}$, define $\bar{R}=\{|r_{ij}|\}$. Then, the following implications hold:
\begin{itemize}
\item[] $R$ is walk-summable 
\item[$\Leftrightarrow$] $\bar{R}$ is walk-summable
\item[$\Leftrightarrow$] $\sum_{\ell} \bar{R}^{\ell}$ converges
\item[$\Leftrightarrow$] $\rho(\bar{R})<1$
\item[$\Rightarrow$] $\sum_{\ell} R^{\ell}$ converges
\item[$\Leftrightarrow$] $\rho(R)<1$ 
\item[$\Rightarrow$] $A=D_A(I-R)$ is nonsingular (where $D_A>0$).
\end{itemize}
In the above,  $\rho(\cdot)$ denotes the spectral radius of a matrix.
\end{lem}
\begin{IEEEproof}
By the basic results of analysis \cite{Rudin} (as pointed out in \cite{Mailoutov}), the unordered sum is well defined if and only if it {\em converges absolutely}, i.e., if and only if $\sum_{w: i\mapsto j} |\phi(w)|$ converges (i.e., the sum is finite and unique,  hence, well defined) for all $i,j$. It is clear that $|\phi(w)|$ corresponds to $\bar{R}$. Hence, $R$ is walk-summable if and only if $\bar{R}$ is walk-summable. It follows from (\ref{eq:walk1}) that $\bar{R}$ is walk-summable if and only if $\sum_{\ell} \bar{R}^{\ell}$ converges, which is equivalent to $\rho(\bar{R})<1$ (a basic stability result \cite{Horn}). Taking any integer $N>0$, we have
\begin{align*}
|(\sum_{\ell=0}^N R^{\ell})_{ij}|&\le \sum_{\ell=0}^N |(R^{\ell})_{ij}| =\sum_{\ell=0}^N |\sum_{i\stackrel{\ell}{\mapsto} j} \phi(w)|\\
&\le \sum_{\ell=0}^N \sum_{i\stackrel{\ell}{\mapsto} j} |\phi(w)|=\sum_{\ell=0}^N (\bar{R}^{\ell})_{ij}
\end{align*}
for all $i,j$. Talking $N\rightarrow \infty$, we know that the convergence of $\sum_{\ell} \bar{R}^{\ell}$ implies that of $\sum_{\ell} R^{\ell}$, which is equivalent to $\rho(R)<1$ (using \cite{Horn} again). Finally, $\rho(R)<1$ obviously implies $A=D_A(I-R)$ is nonsingular. 
\end{IEEEproof}
\begin{rem}\label{rem:walk}
We see in the lemma above a seemingly puzzling phenomenon that the walk-summability of $R$ and that of $\bar{R}$ are equivalent, yet the convergence of $\sum_{\ell} \bar{R}^{\ell}$ and that of $\sum_{\ell} R^{\ell}$ are not equivalent. The reason is that walk-summability requires the unordered sum over all walks from node $i$ to node $j$ to be well defined, whereas $\sum_{\ell} R^{\ell}$ corresponds to a particular ordered sum (i.e., the order of $\ell=1, 2, \ldots$). 
\end{rem}

Next, we show that walk-summable systems include diagonally dominant ones. 
\begin{lem}\label{lem:walk2} 
If $A$ is diagonally dominant \cite{Saad}, then $R=I-D_A^{-1}A$ is walk summable.
\end{lem}
\begin{IEEEproof}
Without loss of generality, assume $a_{ii}=1$ for all $i$. Then, $A$ being diagonally dominant and $a_{ii}=1$ imply that $\gamma_i=\sum_{j\ne i} |a_{ij}| <1$ for every $i$. The well-known Gershgorin circle theorem \cite{Saad} says that every eigenvalue $\lambda$ of $A$ lies in at least one of the Gershgorin discs $D(a_{{ii}},\gamma_{i})$.  Hence, $\rho(\bar{R})<1$, then $R$ is walk-summable. 
\end{IEEEproof}

Finally, we comment on the equivalence between walk-summability and the generalised diagonal dominance. Following \cite{Li}, a matrix $A=\{a_{ij}\}$ is said to be generalised diagonally dominant if there exists a positive diagonal matrix $D=\mathrm{diag}\{d_i\}$ such that $D^{-1}AD$ is diagonally dominant, i.e., 
\begin{align*}
|a_{ii}|d_i> \sum_{j\ne i} |a_{ij}|d_j, \ \ \forall i.
\end{align*}
We will call the system (\ref{Axb}) generalised diagonally dominant if $A$ is generalised diagonally dominant. 
\begin{lem}\label{lem:walk4}
The linear system~\ref{Axb} is walk-summable if and only if it is generalised diagonally dominant.  
\end{lem}
\begin{IEEEproof} 
Consider the matrix $A$ in (\ref{Axb}).  Without loss of generality, assume $a_{ii}=1$ for all $i$. Defining the comparison matrix of $A$ by $\mathcal{M}(A)=\{\alpha_{ij}\}$ with $\alpha_{ii}=|a_{ii}|$ and $\alpha_{ij}=-|a_{ij}|$ for $j\ne i$, it is clear that $A$ is generalised diagonally dominant if and only if $\mathcal{M}(A)$ is so. 
It is known (see, e.g., \cite{Li}) that a matrix $A$ is generalised diagonally dominant if and only if $A$ is an $H$-matrix. Note that a matrix $A$ is called an $H$-matrix if its comparison matrix $\mathcal{M}(A)$ is an $M$-matrix. Recall that an $M$-matrix can be written as $sI-\bar{R}$ with $\bar{R}=\{\bar{r}_{ij}\}, \bar{r}_{ij}\ge0$, and $s\ge \rho(\bar{R})$ \cite{Horn}.  In our case, $a_{ii}=1$ for all $i$ and  $s=1$. We see that $A$ is generalised diagonally dominant if and only if $\rho(\bar{R})<1$, and this is the same as $R$ being walk summable, as shown in Lemma~\ref{lem:walk}, or (\ref{Axb}) being walk-summable.
\end{IEEEproof}

We also note that an iterative algorithm for searching the diagonal matrix $D$ can be found in \cite{Li}.

\end{document}